\documentclass{edbk} % Computer Modern font calls
\usepackage{epsfig}

\normallatexbib
\begin{document}

\articletitle[HBT Studies with E917]      %%  this is the SHORT running title
{HBT Studies with E917 at the AGS: A Status Report}  %% full title of your chapter

\author{Burt Holzman$^{1,6}$}

%% affil, email, and abstract are optional
\affil{Physics Department\\
University of Illinois at Chicago and Argonne National Laboratory\\
Chicago, IL, USA}
\email{bholzm1@uic.edu}
%% affil, email, and abstract are optional
\author{For the E917 collaboration:\\
B.~B.~Back$^1$, R.~R.~Betts$^{1,6}$, H.~C.~Britt$^5$, J.~Chang$^3$,
W.~C.~Chang$^3$, C.~Y.~Chi$^4$,
Y.~Y.~Chu$^2$, J.~Cumming$^2$, J.~C.~Dunlop$^8$, W.~Eldredge$^3$,
S.~Y.~Fung$^3$, R.~Ganz$^{6,9}$,
E.~Garcia-Solis$^7$, A.~Gillitzer$^{1,10}$, G.~Heintzelman$^8$,
W.~Henning$^1$, D.~J.~Hofman$^1$,
B.~Holzman$^{1,6}$, J.~H.~Kang$^{12}$, E.~J.~Kim$^{12}$, S.~Y.~Kim$^{12}$,
Y.~Kwon$^{12}$, D.~McLeod$^6$,
A.~Mignerey$^7$, M.~Moulson$^4$, V.~Nanal$^1$,  C.~A.~Ogilvie$^8$,
R.~Pak$^{11}$, A.~Ruangma$^7$,
D.~Russ$^7$, R.~Seto$^3$, J.~Stanskas$^7$, G.~S.~F.~Stephans$^8$,
H.~Wang$^3$, F.~Wolfs$^{11}$,
A.~H.~Wuosmaa$^1$, H.~Xiang$^3$, G.~Xu$^3$, H.~Yao$^8$, C.~Zou$^3$}

\affil{
$^1$ Argonne National Laboratory, Argonne, IL \\
$^2$ Brookhaven National Laboratory, Upton, NY\\
$^3$ University of California at Riverside, Riverside, CA\\
$^4$ Columbia University, Nevis Laboratories, Irvington, NY\\
$^5$ Department of Energy, Division of Nuclear Physics, Germantown, MD\\
$^6$ University of Illinois at Chicago, Chicago, IL\\
$^7$ University of Maryland, College Park, MD\\
$^8$ Massachusetts Institute of Technology, Cambridge, MA\\
$^9$ Max Planck Institute f\"ur Physik, M\"unchen, Germany\\
$^{10}$ Technische Universit\"at M\"unchen, Garching, Germany\\
$^{11}$ University of Rochester, Rochester, NY\\
$^{12}$ Yonsei University, Seoul, Korea\\
}

%% optional, to supply a shorter version of the title for the running head:
%%\chaptitlerunninghead{}
%\psdraft
\begin{abstract}
Two-particle correlations between pions in Au+Au collisions 
have been measured 
at beam kinetic energies of 6, 8, and 10.8 GeV/u at
the Alternating Gradient Synchrotron (AGS) over a wide range of 
rapidities using a magnetic spectrometer.  The data
have been analyzed in the Hanbury-Brown and Twiss (HBT)
framework to extract source parameters.
The event-by-event orientation of the reaction plane has also been measured 
using a scintillator hodoscope at far forward rapidities, and beam vertexing
detectors upstream of the target.  A preliminary analysis of the dependence
of the source parameters on the reaction plane is
presented.
\end{abstract}

\begin{keywords}
AGS, HBT, reaction plane
\end{keywords}
\section{Introduction}
Identical particle correlations have previously been used to determine
source sizes and lifetimes of the emission regions formed in heavy-ion
collisions \cite{bh_NA49_98,bh_NA44_98,bh_E859_95,bh_e877_97}. In the case of 
non-central
collisions, such measurements usually
integrate over the reaction plane of the collision and
therefore obscure physics which depends on the relative orientation of
the two colliding nuclei.  

In this report we discuss preliminary results
from our investigations of the dependence of
HBT source parameters on the reaction plane.  This approach should
eventually give us another perspective into the source
dynamics.  Together with other HBT dependencies of the
source geometry and dynamics, we 
may thus be able to form a more complete picture of the
nature and evolution of the emission region.

\subsection{The HBT Correlation Formalism}
Quantum statistics require that the amplitudes for
identical particles be added together before interpreting
the square modulus as a probability.  Specifically, identical
bosons, such as pions, are required to have symmetric two-particle amplitudes,
and the HBT correlation arises from the cross-term in the square of
the symmetrized amplitude.

For a pair of identical bosons, emitted by an extended source
$\rho({\bf x})$ with four-momenta ${\bf p}_1$ and 
${\bf p}_2$, detected at space-time 
coordinates ${\bf x}_1$ and ${\bf x}_2$, assuming that the 
bosonic wavefunctions can be described by plane waves,
the symmetrized amplitude of this process is: 
\begin{equation}
\Psi_{12}({\bf x}_1,{\bf x}_2,{\bf p}_1,{\bf p}_2) = 
\frac{1}{\sqrt{2}}(e^{i({\bf p}_1\cdot {\bf x}_1+{\bf p}_2\cdot {\bf x}_2)}
+ e^{i({\bf p}_1\cdot {\bf x}_2+{\bf p}_2\cdot {\bf x}_1)})
\end{equation}

Which leads to the probability of detecting a pair
with relative four-momentum ${\bf q} = {\bf p}_1 - {\bf p}_2$:

\begin{equation}
C_2({\bf q}) \equiv \frac{1}{N({\bf q})}
{\int d^4x_1 d^4x_2~ \rho({\bf x}_1) \rho({\bf x}_2){|\Psi_{12}|}^2}
= \frac{1 + {|\tilde{\rho}({\bf q})|}^2}{N({\bf q})}
\end{equation}

\noindent
where the two-particle correlation function $C_2({\bf q})$ is simply related
to the Fourier transform of the source density.  The normalization
N({\bf q}) is discussed below in Section \ref{bh_normalization}. 

The above
formulation is only strictly valid for completely incoherent
sources. To account for partial coherence effects, as well as 
contamination from long-lived resonances, an empirical 
variable $\lambda$ is added to the 
definition of the correlation function as a coherence scaling 
parameter:

\begin{equation}
C_2({\bf q}) \equiv \frac{1 + \lambda{|\tilde{\rho}({\bf q})|}^2}
{N({\bf q})}
\end{equation}

\subsection{HBT Parameterizations}

In practice, the source $\rho({\bf r})$ has usually been assumed to be
Gaussian in configuration-space.  In this case, the 
momentum-space distribution
is also Gaussian:

\begin{equation}
\rho({\bf r}) \sim e^{\frac{-{|\vec{r}|}^2}{{\bf R}^2}} \Rightarrow
\tilde{\rho}({\bf q}) \sim e^{-{|\vec{q}|{\bf R}}^2} 
\end{equation}

For multi-dimensional HBT analyses, the relative momentum variable 
${\bf q}$ can be expressed in terms of a variety of orthogonal components,
each of which has model-dependent significance \cite{bh_boal90,bh_wu98}.

In the preliminary analysis presented here,
a simple 3-D Cartesian parameterization is chosen: ($Q_x, Q_y, Q_z$), 
with conjugate source parameters ($R_x, R_y, R_z$).
$R_z$ is taken along the beam axis; $R_x$ is lies in the reaction 
plane; and $R_y$ is orthogonal to both.  

In this case, $C_2({\bf q})$ has the following form:
\begin{equation}
C_2({\bf q}) = 
\frac{1}{N({\bf q})}[1 + \lambda{e^{-{(q_xR_x)}^2-{(q_yR_y)}^2-{(q_zR_z)}^2}}]
\label{bh_Geometrical_HBT}
%C_2({\bf q}) = 
%\frac{1 + \lambda{e^{-{(q_xR_x)}^2-{(q_yR_y)}^2-{(q_zR_z)}^2}}}{N({\bf q})}
%\label{Geometrical_HBT}
\end{equation}

\subsection{Reaction Plane Determination}
The reaction plane is determined using the relative orientation of two axes: the 
direction of the incoming
beam particle, $\hat{z}$, and the direction of the impact parameter 
$\hat{b}$.  In our
experiment, 
the beam axis is defined by a beam vertexing detector (BVER).  The
BVER detector consists of four planes of scintillating fibers each read out
by a multi-anode photomultiplier tube.  The fiber planes each consist
of $\sim$ 150 200 x 200 $\mu m^2$ fibers, situated 
5.84 m and 1.72 m 
upstream from the target \cite{bh_back98}. 
The position of the projection of $\hat{z}$ onto the hodoscope can be 
determined with an accuracy of 1.5 mm at 11.4 m downstream from the target. 
%This allows a determination 
%of $\hat{z}$  with an accuracy of 1.5 mm at a position 11.4 m downstream 
%from the target.  

Charged projectile spectator fragments are detected in a 
hodoscope (HODO), and
their charge centroid calculated for each event.  HODO consists of two 
orthogonal planes
of 38 plastic scintillator slats with 1 cm$^2$ cross-sections, 
centered on the 
beam line, and situated 11.4 m downstream from the target. The 
response of individual
scintillators to deposited charge was calibrated on a run-by-run basis and
this information was used to find the charge-weighted centroid 
$\vec{Q} = Q_x\hat{i} + Q_y\hat{j}$ for each event, where
\begin{equation}
Q_x = {\sum Q \cdot x \over \sum Q}
\end{equation}
\label{bh_RP_Defn}
The direction of the impact parameter $\hat{b}$ is then
$\hat{Q}$, defined with the origin at the projected beam position
on HODO, and the reaction plane is defined as the plane spanned by $\hat{Q}$ 
and $\hat{z}$.  $\hat{x}$ is then redefined to lie along $\hat{Q}$.
Implicit in this definition of the impact
parameter is the assumption that the direction
of proton flow -- the deflection of the spectator fragment -- is along
the reaction plane.  

\begin{figure}[ht]
\centerline{
\epsfig{file=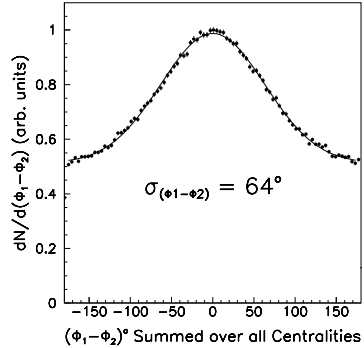,height=1.8in
}}
\caption{Distribution of the reaction plane angle difference for angles
$\phi_1$ and $\phi_2$ determined by splitting each event into two
sub-events.   The angle resolution for the full event is roughly half this
value ($\delta\phi \approx$ 32$^\circ$).}
\label{bh_rp_resolution}
\end{figure}

An estimate of the reaction plane resolution is determined by randomly
dividing each event into two sub-events and looking at the 
($\phi_1~-~\phi_2$) difference distribution 
for the two reaction planes calculated
from each sub-event (see Fig. \ref{bh_rp_resolution}).  The actual resolution 
for the reaction
plane determined using the full event statistics is roughly half this
value \cite{bh_ahle98}, and in this manner we obtain an estimate of
$\delta\phi \approx 32^{\circ}$.
\subsection{Normalization}
\label{bh_normalization}
The correlation function is, by definition, a normalized quantity.
To find the proper normalization N({\bf q}), a background is generated
which creates two-particle ``events'' out of single tracks
from different events.  $C_2({\bf q})$ is then simply
the ratio between real data and the event-mixed background.

\subsection{Corrections to the correlation function}

The shape of the measured two-particle correlation function
mainly depends on three effects: the Coulomb
repulsion between the two particles, the
two-particle resolution of the detector, and the HBT correlation itself.

For the Coulomb effect, a correction 
$f_{C}$ is numerically 
calculated for a simple extended source \cite{bh_pratt86,bh_baker96} of size
$R_0$.  This correction is applied iteratively to the background until
the value of $R_{inv}$ obtained from the one-dimensional correlation function
converges to that of $R_0$.

\begin{figure}[ht]
\centerline{
\epsfig{file=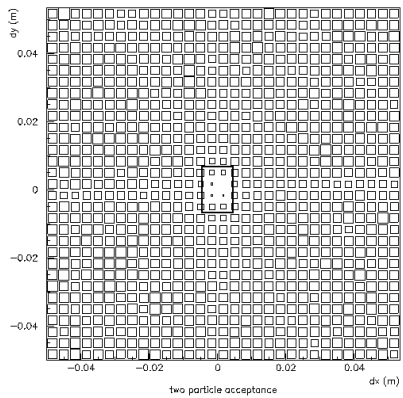,height=150pt
}}
\caption{Separation between two tracks on T1. The $f_{\rm TPR}$ cut is the
solid line.}
\label{bh_TPAC}
\end{figure}

The two-particle resolution arises from the finite resolution of
the tracking detectors in
the spectrometer. Two close tracks are, at some point, indistinguishable 
from single tracks, and do not, therefore, appear in the
measured two-particle correlation.  A two-dimensional cut
in relative coordinate space, $f_{\rm TPR}$, 
is applied to both signal and background events.  The separation between
two tracks, projected onto the first plane in the spectrometer, is shown 
in figure \ref{bh_TPAC}, together with $f_{\rm TPR}$.  

The final correlation function is fitted from a spectrum in {\bf q},
which is generated by dividing signal events by background events.
Both signal and background have been corrected by $f_{\rm TPR}$; the
background has additionally been corrected by $f_C$.

%The final correlation function has this form:
%
%\begin{equation}
%C_2({\bf q}) = \frac{Actual}{Background}
%\frac{f_{TPR}}{f_{TPR}*f_C}
%\end{equation}
%
\section{Results}
The reader is reminded that the following results and analysis
are preliminary, and only contain a limited subset of the E917 data. Thus
far, only 
about 10\% of the data has passed through the HBT analysis.
By December 1999, we expect at least another 50\% will have been analyzed.

\begin{figure}[ht]
%\vskip1.8in
\centerline{
\epsfig{file=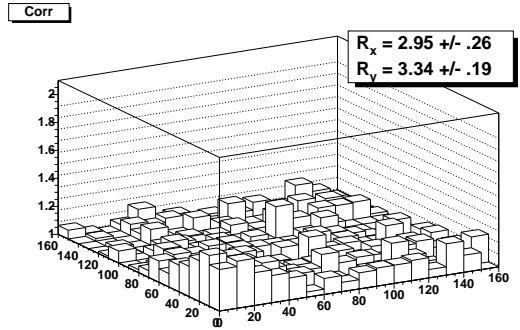,height=1.8in}}
\caption{$|Q_y|$ vs. $|Q_x|$}
\label{bh_2D}
\end{figure}

The measured correlation for identical pairs 
of pions is shown in Fig. \ref{bh_2D}, 
plotted as a function of $Q_x$ and $Q_y$, where x and y are defined
relative to the reaction plane of the event as discussed in Section 
\ref{bh_RP_Defn}.  To improve the statistics, all of $Q_z$ has been 
integrated over.  In addition, since the radii for $\pi^+\pi^+$ and 
$\pi^-\pi^-$ pairs are similar \cite{bh_baker96}, both datasets were 
combined in the present analysis.

\begin{figure}[ht]
%\vskip1.8in
\epsfig{file=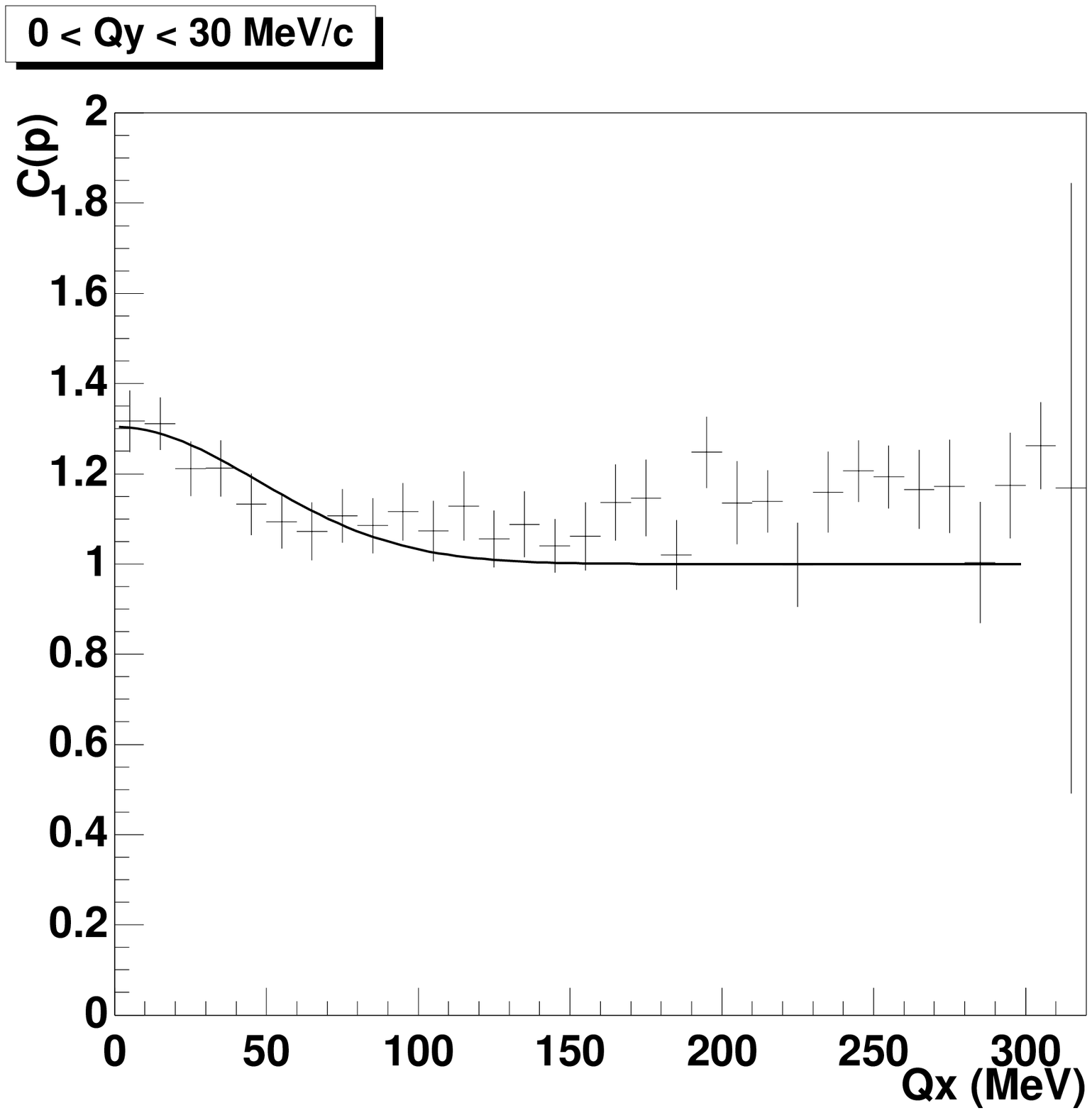,height=1.8in}
\epsfig{file=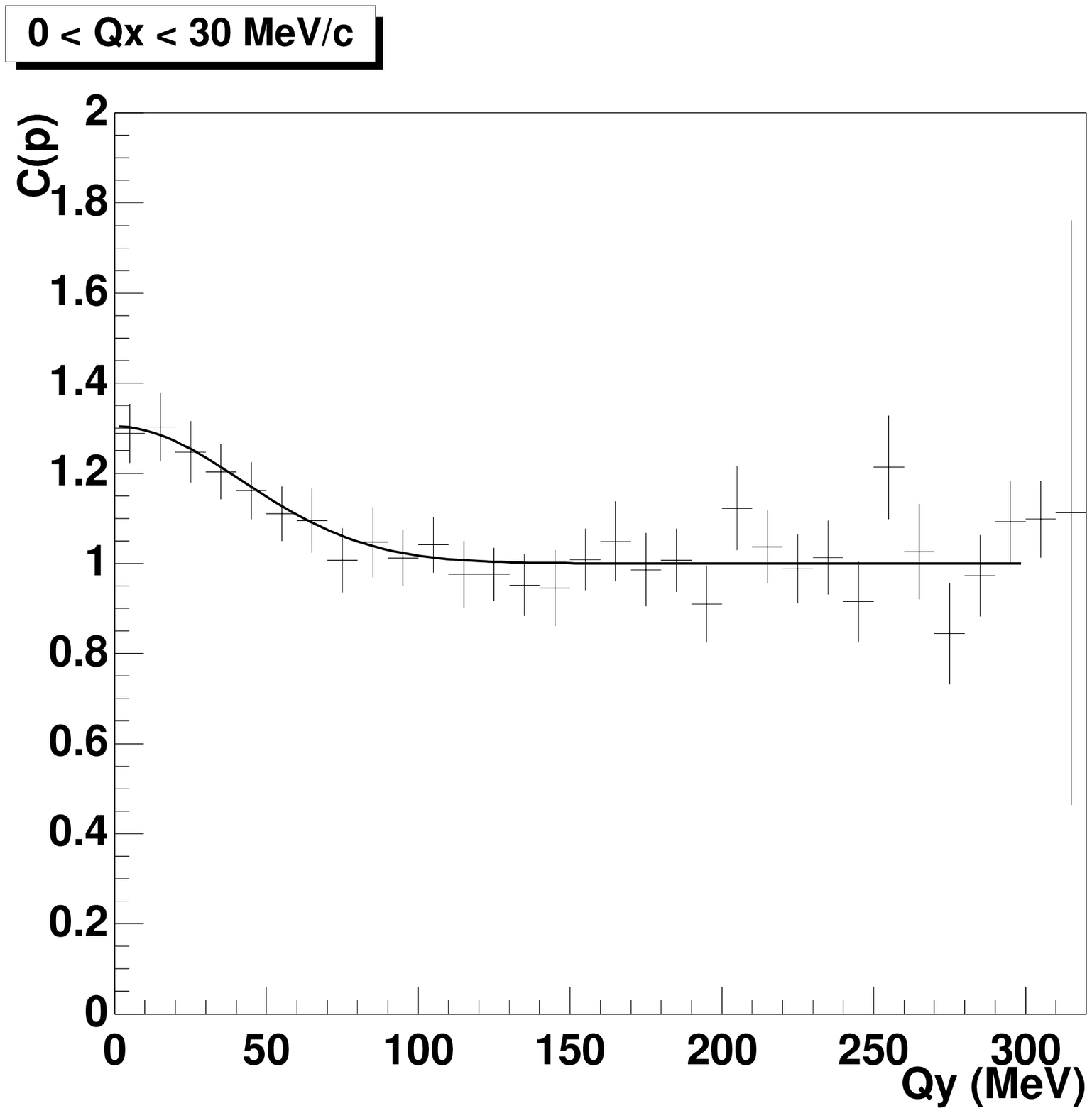,height=1.8in}
\dblcaption{\label{bh_2D_x}$|Q_x|$, $|Q_y| < 30$ MeV}
{\label{bh_2D_y}$|Q_y|$, $|Q_x| < 30$ MeV}
\end{figure}

Central slices are shown in Figures \ref{bh_2D_x} and 
\ref{bh_2D_y} along with the projections from the 
two-dimensional fit.  These data have been fit with the function
in Eq. \ref{bh_Geometrical_HBT}, and values for $R_x$ and $R_y$ obtained.
We find that $R_x = 2.95 \pm 0.26$ fm and $R_y = 3.34 \pm 0.19$ fm, 
the latter value being
larger than the former.  The error bars are solely from the fitting
procedure and do not include systematic errors.

\begin{figure}[ht]
\centerline{
\epsfig{file=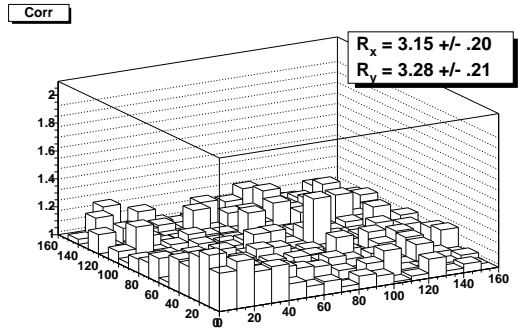,height=1.8in}}
\caption{$|Q_y|$ vs. $|Q_x|$ -- systematic check}
\label{bh_2Dcheck}
\end{figure}

\begin{figure}[ht]
%\vskip1.8in
\epsfig{file=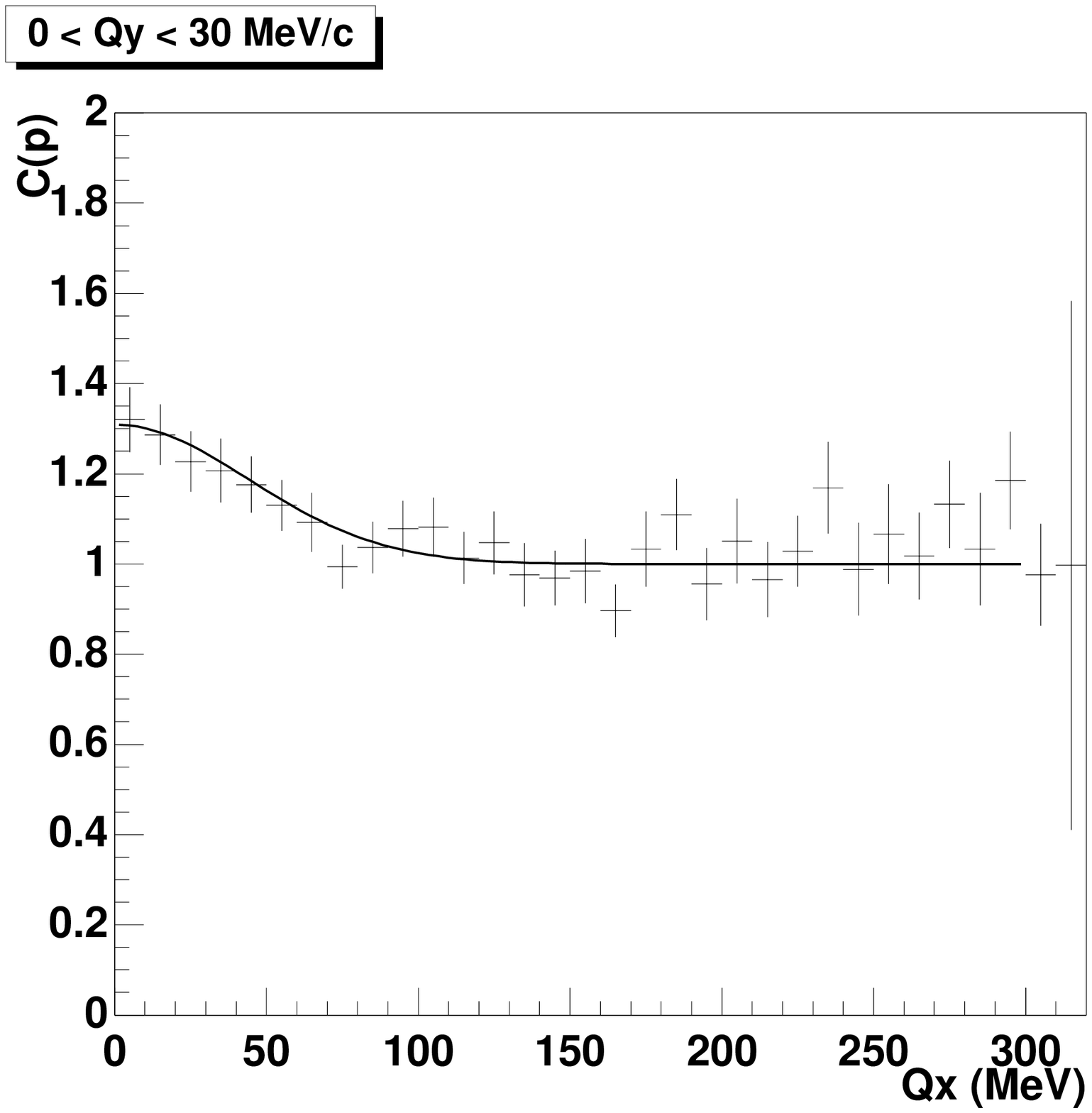,height=1.8in}
\epsfig{file=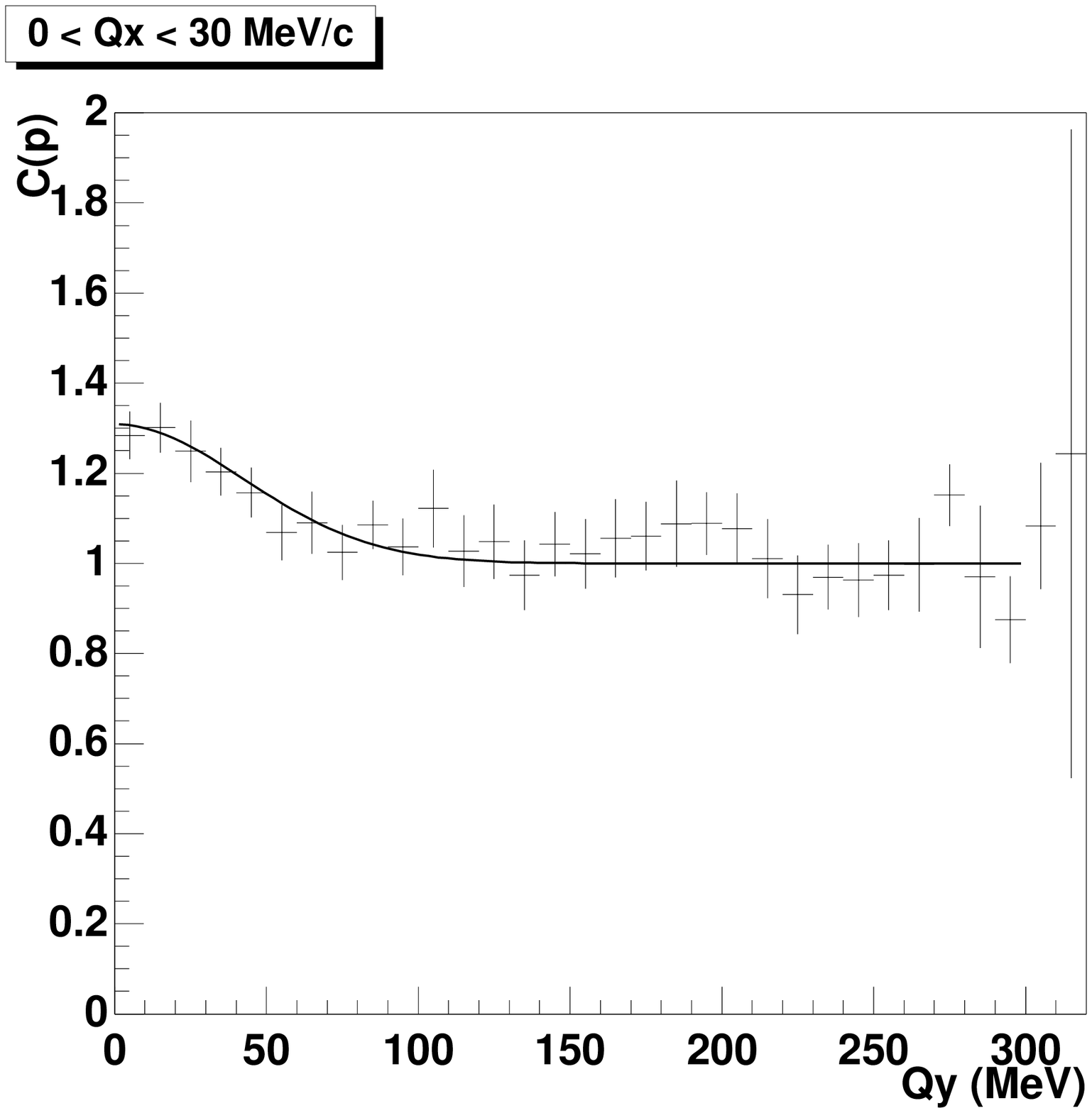,height=1.8in}
\dblcaption
{\label{bh_2Dcheck_x}$|Q_x|$, $|Q_y| < 30$ MeV}
{\label{bh_2Dcheck_y}$|Q_y|$, $|Q_x| < 30$ MeV}
\end{figure}

To establish a baseline for comparison, the data were also
analyzed in the same fashion, but with
$\hat{x}$ chosen along a random direction 
rather than along the reaction plane 
(Figs. \ref{bh_2Dcheck}, \ref{bh_2Dcheck_x}, \ref{bh_2Dcheck_y}).
In this analysis, the values of $R_x$ and $R_y$ obtained are consistent
with each other, indicating that the observed difference relative
to the reaction plane is a real effect.

\section{Conclusions}

In a naive geometrical model of the source, we would expect $R_y$ to
be larger than $R_x$.  This is consistent with the simple overlap of two
non-centrally colliding spheres for which the overlap region resembles an
almond in shape, with $R_y > R_x$.  

Other data and analyses \cite{bh_back99,bh_ollitraut98}, however, 
indicate that the
emission region is not a simple static source at these energies, with
$dN/dy$ distributions and flow studies indicating both 
longitudinal and transverse expansion.  The shape of the source 
may also be obscured by particle absorption by spectator matter.

Future analyses with better statistics will allow the 
investigation of the source
asymmetry with respect to the reaction plane and its dependence on collision 
centrality, pair $m_T$,
and rapidity. These, taken together with $dN/dy$ distributions 
and flow
studies, may be able to further disentangle expansion and
absorption effects 
from the source shape and size, and thus 
assemble a more complete picture of the
collision dynamics.  This work is in progress.

\end{document}